\documentclass[fleqn,usenatbib]{mnras}

\usepackage[T1]{fontenc}

\DeclareRobustCommand{\VAN}[3]{#2}
\let\VANthebibliography\thebibliography
\def\thebibliography{\DeclareRobustCommand{\VAN}[3]{##3}\VANthebibliography}

\usepackage{graphicx}   % Including figure files
\usepackage{amsmath}
\usepackage{natbib} % Advanced maths commands
\title[Modelling UX Ori Stars Eclipses. RR Tau]{Modelling UX Ori Star Eclipses based on Spectral Observations with the Nordic Optical Telescope.
I. RR Tau}

\author[V. P. Grinin et al.]{
V. P. Grinin,$^{1,2}$\thanks{E-mail: vgcrao@mail.ru} L. V.
Tambovtseva$^{1}$, A. A. Djupvik$^{3,4}$, G. Gahm$^{5}$\thanks{Deceased}, T. Grenman$^{6}$\thanks{Deceased},
H. Weber$^{6}$,
\newauthor
H. Bengtsson$^{7}$, H. De Angelis$^{7}$, G. Duszanowicz$^{7}$,
D. Heinonen$^{7}$, G. Holmberg$^{7}$, T. Karlsson$^{7}$,
\newauthor
M. Larsson$^{7}$, J. Warell$^{7}$, T. Wikander$^{7}$
\\
$^{1}$Pulkovo Astronomical Observatory, Russian Academy of
Sciences, Pulkovskoe sh. 65, St. Petersburg, 196140, Russia\\
$^{2}$St. Petersburg State University, Universitetskii pr. 28, St. Petersburg, 198504, Russia\\
$^{3}$Nordic Optical Telescope, Rambla Jos\'{e} Ana Fern\'{a}ndez P\'{e}rez 7, ES-38711 Bre\~{n}a Baja, Spain\\
$^{4}$Department of Physics and Astronomy, Aarhus University, Munkegade 120, DK-8000 Aarhus C, Denmark\\
$^{5}$Stockholm Observatory, AlbaNova University Center, Stockholm University, 106 91 Stockholm, Sweden \\
$^{6}$ Lule\aa \ University of Technology, SE-971 87 Lule\aa \ , Sweden\\
$^{7}$ SAAF, Svensk Amat\"{o}rAstronomisk F\"{o}rening, Sweden
}

\date{Accepted XXX. Received YYY; in original form ZZZ}

\pubyear{2023}

\begin{document}
\label{firstpage}
\pagerange{\pageref{firstpage}--\pageref{lastpage}}
\maketitle

% Abstract of the paper
\begin{abstract}
Based on observations obtained with the Nordic Optical Telescope we
investigate the spectral variability of the Herbig~Ae star RR~Tau.
This star belongs to the UX~Ori family, characterized by very
deep fadings caused by the screening of the star with opaque fragments
(clouds) of the protoplanetary discs.
At the moments of such minima one observes strong spectral variability
due to the fact that the dust cloud occults, for an observer,
not only the star but also a part of the region where the emission
spectrum originates. We calculated
a series of obscuration models to interpret the observed variability of
the H$\alpha$ line parameters. We consider two main obscuration scenarios:
(1) the dust screen rises vertically above the circumstellar disc,
and (2) the screen intersects the line-of-sight moving azimuthally
with the disc. In both cases the model of the emission region consists of
a compact magnetosphere and a magneto-centrifugal disc wind.
Comparison with observations shows that the first scenario explains well
the variability of the radiation flux, the equivalent width,
as well as the asymmetry of the H$\alpha$ line during eclipses, while
the second scenario explains them only partly. This permits us to
suggest that in the case of RR~Tau, the main causes of the eclipses are
either a structured disc wind, or the charged dust lifted along the field
lines of the poloidal component of the magnetic field of the circumstellar
disc.

\end{abstract}

% Select between one and six entries from the list of approved keywords.
% Don't make up new ones.

%%%%%%%%%%%%%%%%%%%%%%%%%%%%%%%%%%%%%%%%%%%%%%%%%%

%%%%%%%%%%%%%%%%% BODY OF PAPER %%%%%%%%%%%%%%%%%%

\section{Introduction}
The UX Orionis type of stars (UXORs) are young stars of mainly A -
F spectral types. They stand out from other young stars by
exhibiting large amplitude flux variations in a specific manner:
their brightness undergoes sporadic weakening with an amplitude of
up to 2-3$^m$ in the V - band and with duration from several days
to several weeks \citep[see e.g.][]{rost97}. The origin of this
variability lies is the non-homogeneous structure of their
protoplanetary discs and the small inclination of the latter to
the line of sight \citep{gri91}, giving rise to changes in the
circumstellar (CS) extinction caused by the passage of opaque gas
and dust fragments from the CS discs across the stellar disc.

During such events the observed radiation of UXORs strongly
changes. Its linear polarization increases. When the stars fade
their reddening turns to blueing although the brightness continues
to decrease (the so-called blueing effect). The scattered
radiation of the protoplanetary discs plays a key role in all
these changes \citep{gri88}. In the bright state of the star the
scattered light is present as a small addition to the observed
radiation of the star. During eclipses its relative contribution
increases because the dusty screen serves as a natural coronograph:
when occulting the star it does not obscure the protoplanetary
disc, as seen from the observer\footnote{Exceptions
are very long-lasting
eclipses when the dust screen is able to obscure the inner part of
the disc, where a noticeable part of the scattered radiation forms
\citep{shg22}.}. Consequently, the brightness of the star cannot
diminish more than to a given level determined by the scattered
radiation. This is why the amplitude of the brightness minima in
all UXORs do not exceed 2 - 3 stellar magnitudes: namely such is
the level of the scattered radiation in the optical region for a
typical protoplanetary disc (about 10\% of the
out-of-eclipse star brightness).

Another consequence of the coronographic effect is the spectral
variability of UXORs during eclipses. When the star fades, the
equivalent width of the H$\alpha$ emission line increases
\citep{ko77,H83,gri94,kgr00,koz06,rod02}.
This is explained by the fact that the emission spectra are formed
in the extended regions which are only partially obscured by the
dusty screen.

The spectrographs used in most of the works devoted to UXORs have
had a low spectral resolution. In this paper we present the first
results of a large program of monitoring of the spectral
variability of these stars, based on high resolution spectra
obtained with the Nordic Optical Telescope (NOT). The star RR~Tau
investigated in the present paper is one of the most
photometrically active UXORs. Using the spectra obtained at the
different states of their brightness, we model the observed
variability of the H$\alpha$ line. We probe physical features and
parameters both of the screening object and the obscured disc
wind.

\section{Observations}

\subsection{Observational strategy}

The observational project aimed at studying the behaviour of the
line profiles as the program stars were going through their faint
states. We selected four different UXORs, among them RR~Tau, the
focus of this paper. In order to sample spectroscopically these
rapidly and irregularly variable stars and follow them through deep
fading events, two requirements were essential: 1) a flexible telescope
scheduling mode and 2) frequent photometric monitoring and a rapid
communication of alerts in order to know when to trigger the
spectroscopic observations.

The first was fulfilled by the Target-of-Opportunity (ToO) program
offered at the NOT \citep{dju10} and its
permanently available fibre-fed spectrograph.
The second has been possible through strong synergy with observers in
the Swedish Amateur Astronomy association SAAF\footnote{Svensk
Amat\"orAstronomisk F\"orening} who reported photometric magnitudes
and alerted when a fading seemed to be taking place. Due to the rapid
and irregular flux variability, it happened that what seemed to be the
start of a fading, was not, or a fading could be over before we could
obtain spectroscopy. Evidently, good weather was required both for the
photometric monitoring as well as for the follow-up spectroscopy, and
in addition to some luck that the alerts coincided with available service
nights at the NOT. The success of the observational project relied
heavily on the collaboration with SAAF.

\subsection{Photometric monitoring by SAAF observers}

\begin{figure*}
  \includegraphics[width=2\columnwidth]{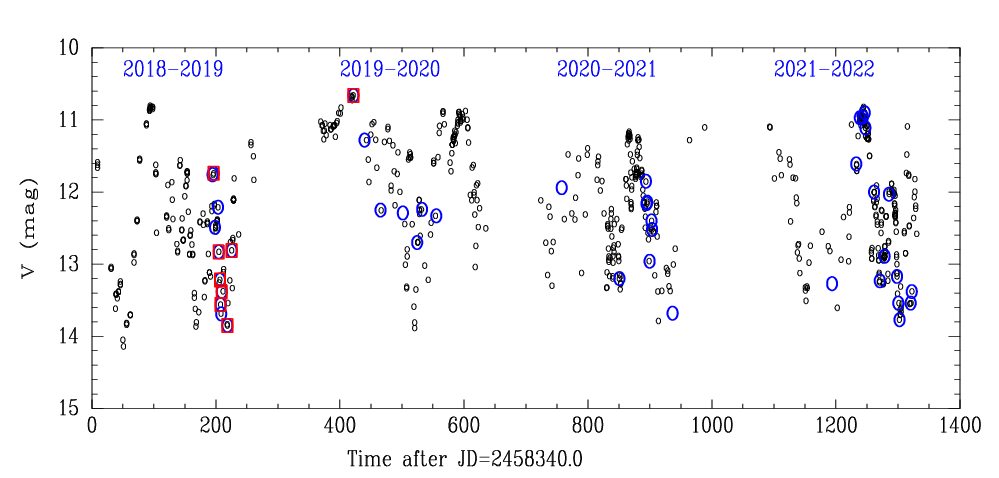}
  \caption{The SAAF V-band photometry of RR~Tau (small black circles) with the epochs of
    FIES/NOT spectroscopy marked (large blue circles). The red squares mark those
    spectra shown in Fig~\ref{fig:obs1}.}
  \label{fig:saaf-vmag}
\end{figure*}

The photometry reported to us by SAAF was used to decide when to trigger a
ToO observation at the NOT. Nevertheless, all the photometric data points
are of value for the study of the photometric variability. These data are
available, since practically all the SAAF photometry was promptly
uploaded to the
AAVSO\footnote{The American Association of Variable Star Observers} database.
An overview of the photometry from SAAF is shown in Fig.~\ref{fig:saaf-vmag}
together with indications of when spectra were obtained.
In this paper we list only the V-band magnitude obtained nearest in time
to each spectrum in Table~\ref{tab:obslist} in order to give an
idea of the faintness of the target at each epoch. In some cases the nearest
photometric data point was obtained from the general AAVSO database. The
separation in time between the photometry and the spectrum is given as an
offset in fractional days. The median time difference is 6 hours.

\subsection{FIES/NOT spectroscopy}
We used the NOT's high-resolution FIber-fed Echelle Spectrograph
(FIES) with its low-resolution fibre, giving a resolving power of
R $\sim$ 25000 \citep{tel14}. We aimed at the above resolution in
order to provide a S/N ratio of around 30 in the V-band continuum
while limiting the exposure time to 1800 seconds even when the
target was in a faint state. As seen in Table~\ref{tab:obslist}
this aim was not always reached, mainly due to adverse sky
conditions. In spite of the sometimes low S/N ratio in the
continuum, however, the emission lines relative to the continuum
increase by contrast when the star fades, therefore providing a
sufficient spectral quality in most cases. The FIES throughput
augmented after recoating the spectrograph mirrors in July 2021.

The wavelength coverage of FIES is around 3700 - 9100~\AA \ in one
setting for the low-resolution fibre. (The exact wavelength
coverage shifted slightly upon operation inside the spectrograph,
for instance when installing a pressure tank for the echellogram
in June 2019.) The more than 90 spectral orders are spread out on
the 2k $\times$ 2k detector (CCD15), a deep-depleted CCD from e2v
installed in 2016. Beyond 8300~\AA \ there is no order overlap,
but some gaps between orders, because the detector is not big
enough to sample the full echellogram. Standard calibrations (21
halogen flats, 7 biases, and one ThAr) were obtained each
afternoon. Data reduction was made with the FIEStool
pipeline\footnote{FIEStool manual, Stempels 2005, Nordic Optical
Telescope, see \url{http://www.not.iac.es/instruments/fies/}}
version 1.5.1 until August 2021 when upgraded to version 1.5.2
with improved handling of the merging of overlapping orders. The
merged spectra output from FIEStool were normalised by fitting a
continuum and put on a heliocentric velocity grid using standard
tasks in the IRAF package.

A specific feature seen in the RR~Tau spectra in deep minima are ghosts from
the very strong H$\alpha$ emission line which peaks at 40 times the continuum.
The ghosts are visible as broad double-peaked emission profiles in the
neighbouring orders.

\begin{table}
  \caption{Log of FIES/NOT observations of RR~Tau showing each epoch as dates
    and UT time of the midtime of the exposure, where $\chi$ is the airmass,
    S/N is the measured S/N ratio around 5300 \AA, $V$ is the nearest in time V-band photometry
    and $\sigma_V$ its error, and $\Delta$t gives the time difference between
    spectroscopy and photometry in fractional days (negative value
    means photometry was obtained before the spectroscopy).}
  \label{tab:obslist}
  % Table rrtau_tab.tab  Wed 22:58:14 12-Apr-2023
%\documentstyle{article}
%\begin{document}
\newcommand\cola {\null}
\newcommand\colb {&}
\newcommand\colc {&}
\newcommand\cold {&}
\newcommand\cole {&}
\newcommand\colf {&}
\newcommand\eol{\\}
\newcommand\extline{&&&&&\eol}

\begin{tabular}{lcrccr}
\hline
\cola DATE-TIME\colb $\chi$\colc S/N\cold  V\cole $\sigma_V$\colf $\Delta$t\eol
\cola          \colb     \colc    \cold mag\cole     mag\colf  days\eol
\hline
\cola 2019-02-19T22:59:30.86\colb 1.10\colc  23\cold 11.76\cole  0.03\colf  1.04\eol
\cola 2019-02-21T20:36:06.43\colb 1.01\colc  38\cold 11.74\cole  0.02\colf -0.07\eol
\cola 2019-02-23T22:21:30.47\colb 1.06\colc  26\cold 12.49\cole  0.01\colf  0.87\eol
\cola 2019-03-01T00:44:39.97\colb 1.74\colc  31\cold 12.21\cole  0.04\colf -0.03\eol
\cola 2019-03-01T21:30:28.18\colb 1.03\colc  27\cold 12.83\cole  0.05\colf  0.28\eol
\cola 2019-03-03T21:53:49.98\colb 1.07\colc  15\cold 13.22\cole  0.04\colf  0.10\eol
\cola 2019-03-04T21:44:01.71\colb 1.06\colc  18\cold 13.56\cole  0.05\colf  0.05\eol
\cola 2019-03-05T22:13:37.97\colb 1.12\colc   5\cold 13.69\cole  0.02\colf -0.14\eol
\cola 2019-03-06T22:37:21.18\colb 1.19\colc  13\cold 13.38\cole  0.03\colf  2.10\eol
\cola 2019-03-15T22:32:47.64\colb 1.29\colc  11\cold 13.85\cole  0.05\colf  0.03\eol
\cola 2019-03-22T21:30:34.36\colb 1.17\colc  28\cold 12.81\cole  0.04\colf  0.04\eol
\cola 2019-10-05T05:44:09.37\colb 1.01\colc  81\cold 10.66\cole  0.03\colf -0.13\eol
\cola 2019-10-23T06:16:50.18\colb 1.04\colc  48\cold 11.28\cole  0.02\colf  2.87\eol
\cola 2019-11-18T02:26:59.04\colb 1.02\colc  26\cold 12.25\cole  0.01\colf  0.79\eol
\cola 2019-12-24T02:49:04.77\colb 1.10\colc  26\cold 12.29\cole  0.10\colf  0.20\eol
\cola 2020-01-16T00:25:32.93\colb 1.03\colc  15\cold 12.70\cole  0.03\colf  0.88\eol
\cola 2020-01-23T21:31:25.25\colb 1.06\colc  14\cold 12.24\cole  0.03\colf -1.02\eol
\cola 2020-02-15T23:01:18.91\colb 1.07\colc  23\cold 12.33\cole  0.02\colf -1.10\eol
\cola 2020-09-05T05:20:19.37\colb 1.25\colc  36\cold 11.94\cole  0.10\colf -0.79\eol
\cola 2020-12-07T03:35:39.87\colb 1.07\colc  13\cold 13.20\cole  0.01\colf -0.11\eol
\cola 2021-01-19T00:16:57.89\colb 1.03\colc  32\cold 11.85\cole  0.01\colf -0.09\eol
\cola 2021-01-19T22:08:55.65\colb 1.03\colc  38\cold 12.16\cole  0.06\colf  1.09\eol
\cola 2021-01-20T23:16:07.67\colb 1.00\colc  35\cold 12.14\cole  0.07\colf  0.17\eol
\cola 2021-01-24T23:22:42.22\colb 1.01\colc  22\cold 12.96\cole  0.04\colf -0.73\eol
\cola 2021-01-27T23:28:32.99\colb 1.02\colc  30\cold 12.40\cole  0.01\colf  0.26\eol
\cola 2021-01-29T00:42:48.85\colb 1.14\colc  30\cold 12.52\cole  0.01\colf -0.15\eol
\cola 2021-03-02T23:22:00.51\colb 1.29\colc  12\cold 13.68\cole  0.10\colf -0.88\eol
\cola 2021-11-15T03:06:35.64\colb 1.00\colc  24\cold 13.27\cole  0.10\colf -0.91\eol
\cola 2021-12-24T04:40:59.71\colb 1.54\colc  48\cold 11.61\cole  0.03\colf  0.59\eol
\cola 2021-12-30T00:55:16.85\colb 1.00\colc  57\cold 10.97\cole  0.05\colf -0.03\eol
\cola 2022-01-04T00:30:33.85\colb 1.00\colc  67\cold 10.96\cole  0.03\colf  0.26\eol
\cola 2022-01-07T00:27:09.24\colb 1.01\colc  47\cold 10.90\cole  0.05\colf -0.07\eol
\cola 2022-01-08T02:33:45.72\colb 1.21\colc  52\cold 11.11\cole  0.10\colf -0.10\eol
\cola 2022-01-21T20:57:24.34\colb 1.13\colc  47\cold 12.00\cole  0.03\colf  0.28\eol
\cola 2022-01-31T02:56:47.39\colb 1.95\colc  17\cold 13.23\cole  0.02\colf -0.33\eol
\cola 2022-02-07T22:37:24.84\colb 1.01\colc  28\cold 12.89\cole  0.01\colf -0.13\eol
\cola 2022-02-14T21:16:51.39\colb 1.00\colc  41\cold 12.03\cole  0.03\colf  0.34\eol
\cola 2022-02-27T22:56:20.11\colb 1.17\colc  21\cold 13.17\cole  0.10\colf -0.08\eol
\cola 2022-03-01T21:14:07.14\colb 1.02\colc  18\cold 13.54\cole  0.05\colf  0.10\eol
\cola 2022-03-03T22:09:27.26\colb 1.10\colc  13\cold 13.77\cole  0.05\colf -0.08\eol
\cola 2022-03-21T21:42:39.13\colb 1.20\colc  10\cold 13.54\cole  0.03\colf -1.14\eol
\cola 2022-03-23T20:45:12.14\colb 1.09\colc  16\cold 13.38\cole  0.04\colf  0.24\eol
\cola 2022-10-05T04:24:48.75\colb 1.09\colc   6\cold n.a.\cole   n.a.\colf  n.a.\eol
\cola 2022-10-08T01:42:32.91\colb 1.81\colc  48\cold 11.30\cole  0.10\colf  0.96\eol
\cola 2023-01-06T23:03:46.17\colb 1.03\colc  63\cold 10.80\cole  0.10\colf -4.12\eol
\hline
\end{tabular}

%\end{document}

\end{table}

\section{Observational data}

For RR~Tau we obtained 45 spectra in the period from February 2019 to
January 2023, spanning a brightness range from V = 10.7 mag to V = 13.9 mag.
The observational details are listed in Table~\ref{tab:obslist}.

\begin{figure*}
\centering
\includegraphics[width=40mm]{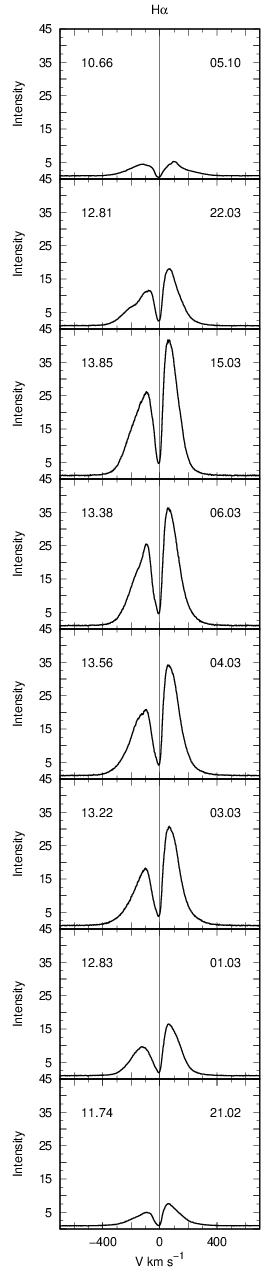}
\includegraphics[width=40mm]{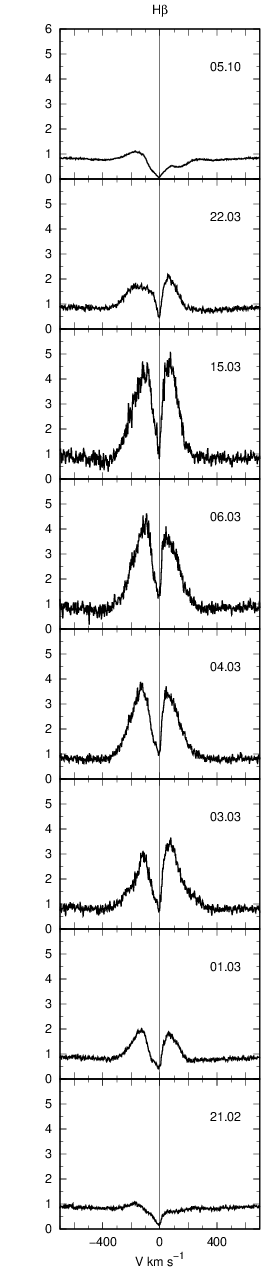}
\includegraphics[width=40mm]{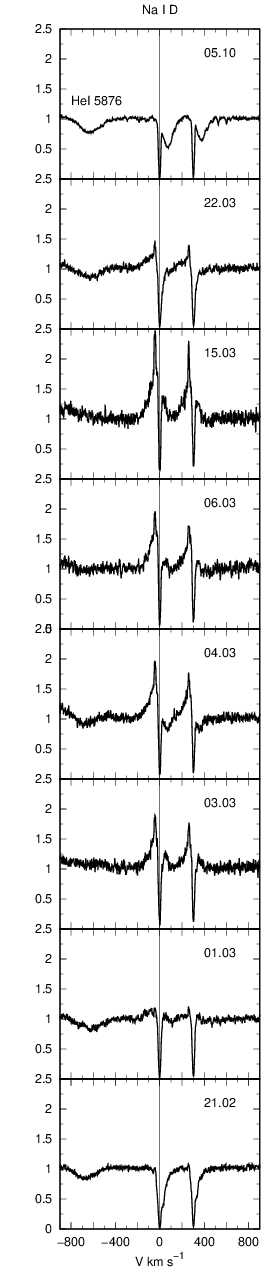}
\includegraphics[width=40mm]{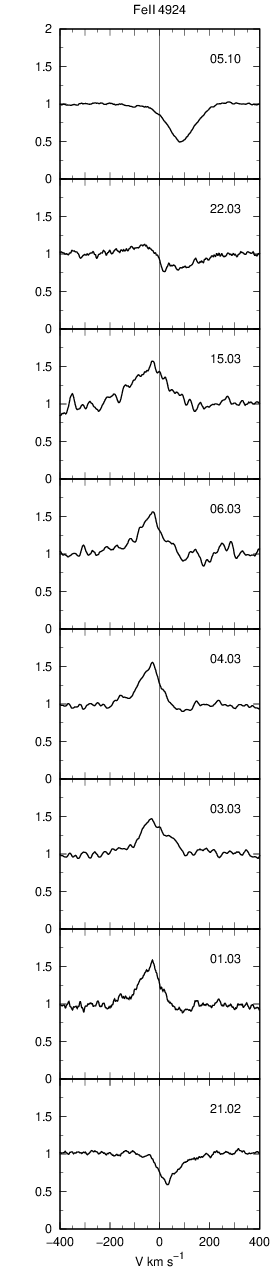}

\caption{H$\alpha$, H$\beta$, Na~I~D, and Fe~II~4924 \AA \ line profiles
  at the maxima and minima brightness stages in 2019. The dates of the
  observations are shown in each plot, and the left panel includes the
  value of the V-band magnitude. Details are in the text.} \label{fig:obs1}
\end{figure*}

Examples of selected regions of some observed spectra are shown in
Fig.~\ref{fig:obs1} in the star coordinate system, using the radial velocity
RV~=~11~km~s$^{-1}$ for RR~Tau \citep{grik01}. We chose to show
the eclipse moments observed in March 2019. They are shown between
two bright states of the star (2019-02-21 and 2019-10-05). The
regions are: the prominent hydrogen emission lines H$\alpha$
and H$\beta$, the sodium lines Na~I~D and the photospheric iron line
Fe~II 4924 \AA. The left panel showing H$\alpha$ includes the value of
  the V-magnitude for the given date.
It is seen from the figure that when the star is occulted,
the Balmer hydrogen lines become strongly intensive, the sodium and iron
lines have an emission, and the absorption component of the He~I 5876 \AA \
line weakens or is completely absent (seen next to the sodium lines).
The Fe~II line is observed as a photospheric line in the bright state and
transforms to a blue-shifted emission line in the weak state. Analogous
changes are observed for the two other components of this multiplet. The same
effect was observed in the RR~Tau spectra by \citet{rod02}.

\section{Modelling}
We model the eclipses of RR~Tau in the framework of two obscuration scenarios.
We consider the dust and gas fragment of the protoplanetary disc or/and disc
wind (cloud) which intersects the line-of sight (LOS), and model these
dust clouds as simple screens. Up to now the origin of such fragments
is unknown (see Discussion). Therefore we consider a simplified
model of the screening body. Our goal is to clarify what type of screen motion
is realized or dominates at the moments of the eclipse: vertical or horizontal.

\subsection{Obscuration scenarios}
\begin{figure}
\begin{centering}
\hspace{0.8cm}\includegraphics[width=7.5cm]{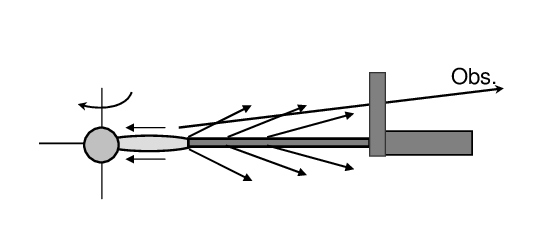}
\includegraphics[width=8cm]{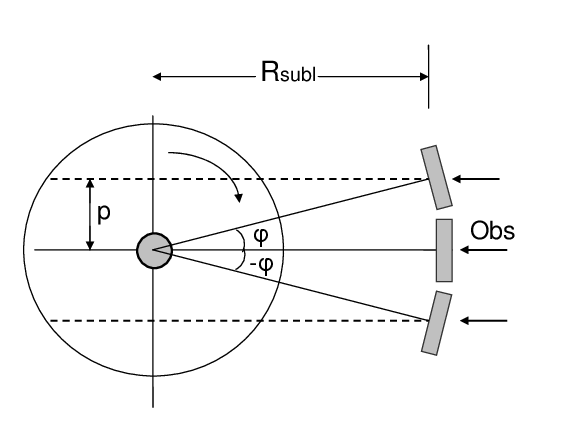}

\caption{Positions of the screen relatively the system "star plus
disc". $\phi=0$ is "the central eclipse". View of the system
edge-on (top) and face-on (bottom). Not to scale.}
\label{fig:ske}
\end{centering}

\end{figure}

In both scenarios the screen is placed at a distance equal to the dust
sublimation radius which is assumed to be at 0.45~AU or 46$R_*$
  \citep{nat01,ddn01,tan07,flo17}.
The edge-on and pole-on sketches of the scenarios are shown in
Fig.~\ref{fig:ske}.

\begin{figure}
\begin{centering}
\includegraphics[width=\columnwidth]{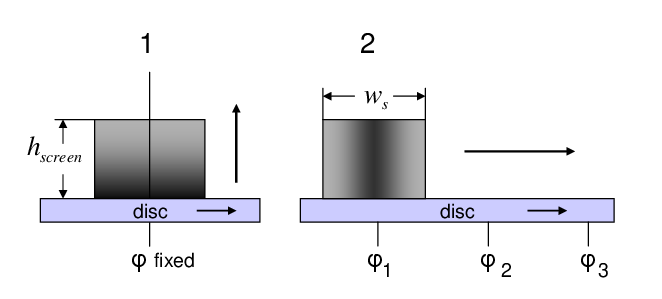}
 \caption{(1): In the first obscuration scenario the screen is fixed at the azimuthal
angle $\phi$ and rises above the disc surface; (2): In the second
scenario the screen of a constant height changes its position moving with the disc.}
\label{fig:ske2}
\end{centering}
\end{figure}

\textit{In the first scenario} we fix the screen at some angle
$\phi$ and raise it up gradually, calculating the line profiles at
each height of the screen (Fig.~\ref{fig:ske2} (1)). We assume that during
an eclipse the location of the screen does not change azimuthally.
Then we reposition the screen to the next azimuthal angle $\phi$
and repeat this procedure. In this manner we "reproduce" a
situation where dust is lifted with the gas in the disc or in the
dusty disc wind. The opacity of the screen does not change over
the screen's width $w_s$ but it varies with its height: the higher
the screen, the more dense layers appear.

The optical depth of the screen is derived as follows
   \begin{equation}
{    \tau_{h}=\tau_0 \cdot \exp(\frac{\Delta h}{b})^{2} },
\end{equation}
where $\Delta h = h_{screen}-h_{cross}$, that is the height of the
screen minus the height counted from the disc surface till the
cross-point of the LOS with the screen, $\tau_h$ is the opacity of
the screen at the height $\Delta h$, and $\tau_0$ and $b$ are free
parameters, chosen in such a way as to avoid a steep opacity
gradient.

\textit{In the second scenario} the screen also has a finite size
$w_s$, and is located at a fixed $\phi$ at the dust
sublimation radius. The height of the screen is fixed and
constant; it has to be high enough to occult the star's light at
the given inclination angle. The optical depth of the screen's
matter does not depend on the screen's height but depends on its
width $w_s$ (Fig.~\ref{fig:ske2} (2)). In the 2nd scenario we "move" the screen
over $\phi$ "reproducing" a situation when a dust and gas cloud -
i.e. a disc or disc-wind fragment, or a protoplanet - orbits the
star and its vicinity. We used the Gauss function to give the
screen's opacity in this case
   \begin{equation}
{    \tau_{ws}=\tau_0 \cdot \exp(-\frac{\Delta w_s}{\sigma})^{2}},
\end{equation}
where $\Delta w_s=w_s/2-w_{cross}$, and $w_{cross}$ is the cross-point
of the screen with the LOS, $\tau_0$ and $\sigma$ are parameters.

\subsection{Disc wind, accretion and scattered light models}

The UX Ori type star RR~Tau is a Herbig~Ae star. Herbig stars do
not possess strong magnetic fields, typical values are 100-200 G
\citep{wade07,wade09,al13a,al13b,hub15,jar19,men20}.
They are also known to be rapidly (100-250 km s$^{-1}$) rotating stars
\citep{mora01,grik01}. One more characteristic feature of these
stars is a compact accreting region that may be a consequence of
the high rotation velocities and weak magnetic fields. This
follows both from observational investigations \citep{CJK14,CJK15}
and from spectroscopic and interferometric modelling of their emission
spectra \citep{gr15,car15,kre18,tam14,tam16,tam20}.
More information on the current knowledge of Herbig stars (and
UXORs) can be found in the recent review by \citet{bri23}.
In the models of the RR~Tau emission spectra we take into
account the contribution of the continuum emission (star) and the
line emission (disc wind plus magnetospheric accretion + disc
scattered radiation).

\subsubsection{Disc wind}
Our disc wind model is based on the theory of a
magneto-centrifugal disc wind developed for the accretion discs
surrounding black holes by \citet{bla82} and applied to young
stellar objects by \citet{pud86}. The geometry and kinematics of
the disc wind obtained from solutions of the magnetohydrodynamic
(MHD) equations can be reproduced with the method of
parametrization. Following \citet{kur06} who developed it for the
low mass T~Tauri stars (TTS), we applied it for Herbig~AeBe stars
taking into account their special features. Besides we introduced
the co-rotation zone because the disc wind removes an excess of
the angular momentum from the system as is shown by the solution
of the MHD equations \citep{saf93}.

We took into account that the gas in the disc wind is rapidly
heated by ambipolar diffusion up to $\sim$ 10 000 K
\citep{saf93,gar01}. We assume that the disc wind consists of
hydrogen atoms with a constant temperature (10 000 K) except for
the regions just above the disc surface where the temperature is
not high enough to excite hydrogen line emission. According to our
calculations the gas has to be heated to at least 6000 K,
therefore we exclude from consideration the very base of the disc
wind.

An algorithm of disc wind modelling is as follows (\citep{tam14});

1) We divide the wind region in a number of streamlines, and choose
the parameters of the wind that determine the needed geometry,
kinematics and mass loss rate. Then we compute the velocity and
density distribution throughout the region.

The main model parameters are as follows. The footpoints of the
disc $w_1$ and $w_N$ refer to the wind launching region for the
first and last streamlines, respectively. $\theta_1$ is the half
opening angle between the first streamline and the vertical axis.
The poloidal velocity component $v(l)$ changes along the
streamlines, as given by
\begin{equation}\label{v}
    v(l) = v_0 + (v_{\infty} - v_0)\,(1 - l_i/l)^{\beta}\,.
\end{equation}
Here $v_0$ and $v_{\infty}$ are the initial and terminal
velocities, and $\beta$ is a free parameter which permits us to
vary the acceleration of the gas. We assume $v_0$ to be the sound
velocity in the disc wind. The terminal velocity is
$v_\infty$\,=$fu_K(w_i)$, where $u_K(w_i)$ is the Keplerian
velocity at the footpoint $w_i$ on the disc surface for the $i$th
streamline, and $f$ is a scale factor. \citet{kur06} used for TTS
$f$ = constant in their calculations. In reality, the parameter
$f$ can change with the distance because the strength of the
magnetic field decreases with $w$.

The tangential velocity $u(w)$ changes with the cylindric radius $w$
as follows: In the co-rotation zone ($u(w)/u_K(w_i)) \leq f_c)$
\begin{equation}\label{u1}
    u(w)=u_K(w_i)(w/w_i);
\end{equation}
in the zone of the conservation of the angular momentum
\begin{equation}\label{u2}
    u(w)=u_K(w_i) f_c/(w/w_i),
\end{equation}
where $f_c$ is a free parameter. It permits us to introduce
a boundary between the corotation zone and the zone of the conservation
of the angular momentum.

The last two parameters are the mass loss rate $\dot M_{w}$ and a parameter
$\gamma$.
\begin{equation}
\dot M_{w} = 2\int\limits_{w_1}^{w_N}\dot m_{w}(w)\,2\,\pi\,w,dw,
\end{equation}
where $\dot m_w$ is the local meaning of the mass loss
rate per unit area on the disc surface and the factor 2 takes into
account the mass loss through both surfaces of the disc (above and
below). The disc itself treated as geometrically thin.
\begin{equation}\dot m_{w}(w) \sim w^{-\gamma}\,,
\end{equation}
    where $\gamma$ is a free parameter which regulates the mass
    loading among the streamlines. The dependence of the line
profiles on $\gamma$ is presented in \citet{gri11}.

2) For all emitting volumes around the star we performed a non-LTE
modelling of the radiative transfer. The algorithm of the
calculations is described in detail by \citet{gri11}. The solution
of the radiative transfer problem is based on the numerical code
developed for media with large velocity gradients. The source
functions were calculated in the Sobolev approximation
\citep{sob60}, and the intensity of the radiation emergent at
frequencies within a spectral line was calculated by exact
integration over spatial coordinates in the approximation of full
redistribution over frequency in the co-moving coordinate system.
We took into consideration 15
hydrogen levels and the continuum, and used the Doppler profile of
the absorption coefficient. The Stark broadening is negligible in
our models.

3) We compute the intensities of the hydrogen line at the given
frequencies in the integration region. The results are presented
as line profiles, all of them are normalized to the star continuum
which varies during an eclipse.

\subsubsection{Accretion}
In spite of the fact that Herbig Ae stars have small magnetic
fields and small accreting region they are able to accrete disc
matter onto the stars \citep[see e.g.][]{vink02,vink05,mot07,aba17}.
Taking into account weak magnetic fields and fast rotation (100 -
200 km s$^{-1}$), one can expect a different configuration of the
magnetosphere compared to that of TTS. We can assume that the
accreting matter does not come to the pole region of the star but
to the middle latitudes, closer to the equator. We call this
region the disc accretion zone.

The radial velocity of the gas $v(r)$ is determined from the
solution of the equation of motion under gravity:
\begin{equation}\label{vms}
    v(r)\frac{dv}{dr} = -\frac{GM_*}{r^2} + \frac{u^2(r)}{r},
\end{equation}
where $r$ is the distance from the stellar center, $G$ the
gravitational constant, and $M_*$ the mass of the star.

The rotational gas velocity $u(r)$ is determined by a power law
\begin{equation}\label{urot}
    u(r)=U_*(r/R_*)^{pm},
\end{equation}
where $r$ is the distance from the star, $U_*$ is the rotational
velocity of the star at the point where gas reaches its surface,
and $pm$ is a parameter. The electron temperature $T_e(r)$ was chosen
following the law
\begin{equation}\label{tm}
    T_e(r) = T_e(R_*) (r/R_*)^{-\alpha}
\end{equation}
where $\alpha$ is a parameter and $T_e(R_*)$ the electron
temperature at the stellar surface.

In the Appendix (Sect.~\ref{appendix-a}) we present the contribution
of the disc wind and accretion region to the total emission in the
H$\alpha$ line (Fig.~\ref{rm4}).

\subsubsection{Scattered light}
As mentioned above, the brightness of UXORs cannot drop during
an eclipse more than a few stellar magnitudes. This is due to the
presence of the stellar radiation scattered by the disc and/or the
disc wind $I_{sc}$. This additional emission can easily be
extracted from the observed total emission $I_{obs}$:
\begin{equation}\label{iobs}
    I_{obs} = I_*\cdot \exp(-\tau_\lambda) + I_{sc}.
\end{equation}
where $I_*$ is the out-of-eclipse radiation of the star,
$\tau_\lambda$ the optical depth of the body screening the star at
the wavelength $\lambda$.

Knowing the maximum changes in the brightness at the moment of an
eclipse from photometric observations $\Delta m_{max}$, one can
obtain the contribution of the stellar radiation scattered by the
CS dust into the out-of-eclipse radiation of the star
\begin{equation}\label{mag}
    (\Delta m)_{max} = 2.5\log(1 + I_*/I_{sc}).
\end{equation}
$\Delta m_{max}$ observed in the minima of UX~Ori itself is
usually 2 - 3 stellar magnitudes while that of RR~Tau is 3 - 4
stellar magnitudes. Thus, estimates of the contribution of the
scattered light for UX~Ori is about 10\% of the out-of-eclipse
star brightness while for RR~Tau this addition is about of 3\%
\citep{rost97}. When modeling we add this amount of the scattered
radiation to the intensity of radiation during eclipses.

\subsubsection{Radiation flux in the H$\alpha$ line} The radiation
flux in the continuum $F_i$ at the observational moment $i$ is
determined as follows
\begin{equation}\label{e12}
    F_i=10^{-0.4 \cdot \Delta V},
\end{equation}
where $\Delta V = V_i-V_0$, and $V_0$ is the star's magnitude
taken in the bright state and used as a unit\footnote{Here we
assume that variation of the flux in the continuum of the star
near the H$\alpha$ line is close to the variation of the flux in
the V-band}. The radiation flux in the H$\alpha$ line is given by
the next expression
\begin{equation}\label{e13}
    F_{\alpha i}= EW_i \cdot F_i ,
\end{equation}
where $EW_i$ is the equivalent width of the H$\alpha$ line at
the same moment $i$. In order to obtain the radiation flux normalized
to the flux in the bright state one has to divide $F_{\alpha i}$
by the meaning of the equivalent width at the bright state $EW_0$.
The radiation flux in the emission lines has to decrease with the
fading of the star because the gas and dust screen obscures
regions of the disc wind.

\section{Result}
The aim of our modelling is to investigate the behavior of the
H$\alpha$ line profiles and their main parameters during eclipses.
As mentioned above, the close vicinity of the star undergoes
strong variations both due to intrinsic reasons (e.g., variable
accretion rate) and external ones (variable extinction). Therefore
we chose accretion and disc wind parameters which reproduce in
total the observed line profiles and do not contradict the
theoretical background.

We considered several hybrid models consisting of the same
accretion model and different disc wind models. Here we present
two of them: H1 and H2 whose disc wind parameters DW1 and DW2
correspondingly are given in Table~\ref{dwtab}.

\begin{table}
 \caption{Disc wind model parameters.}
 \centering
\begin{tabular}{|c|c|c|c|c|c|}
\hline
 Model & $\omega_1 - \omega_N$ & $\theta_1-\theta_N$ & $\dot{M_w}$& $\gamma$ & $\beta$ \\
  \hline
  &$R_*$ & degree & $M_\odot$ yr$^{-1}$ & &  \\
  \hline
  DW1 & 2 - 3.5  & 30 - 45 & $3 \times 10^{-9}$ & 3& 5 \\
\hline
 DW2 & 2 - 6  & 30 - 60 & $5 \times 10^{-9}$ & 3& 5 \\
\hline
\end{tabular}
\label{dwtab}
\end{table}

Here $\omega_1 - \omega_N$ are the cylindric radii restricting the
disc wind launching region, $\theta_1-\theta_N$ are the first and
the last half opening angles of the disc wind, $\gamma$ is a
parameter that "distributes" the matter over streamlines, $\beta$
is a parameter in the poloidal velocity law, and $\dot{M_w}$ is
the total mass loss rate.

In the accretion model $T_e(R_*)$ = 10 000~K, $\alpha$= 1/3, $U_*$ =
70 km s$^{-1}$, the half-thickness of the magnetosphere is
0.75$R_*$, the corotation radius is $1.5R_*$ and the accretion
rate is $10^{-7}M_\odot$ yr$^{-1}$.

We assume next the parameters of RR~Tau to be: the mass of the
star $M_* = 2.5 M_\odot$, the radius R$_* = 2.1 R_\odot, T_{\rm
eff}=9750$ K, $\log g =3.5$ \citep{rost99}. We use the Kurucz
model with $T_{\rm eff} = 10000$~K and $\log g = 3.5$
\citep{kur79}. All calculations have been made for the inclination
angle $i = 70^\circ$ ($ i = 0$ means the pole-on viewing). This
angle is chosen as the more probable for UXORs seen nearly edge-on
\citep[see e.g.][]{gri91,kre13,kre16}.

\subsection{The first obscuration scenario}

The dependence of the optical depth $\tau_{h}$ on the screen's height
obeys Eq. 1. After probing different variants we chose as the appropriate
parameters $\tau_0 = 0.1, b=5$. For both hybrid models, differing only by the
disc wind models, we calculated the H$\alpha$ line profiles with screens of
different widths. Besides we calculated the equivalent widths of the line
profiles, the radiation fluxes in the line $F_{\alpha}$, and the blue-to-red
peak ratios, which following tradition we give as V/R for violet/red.
Thus, we obtained the behavior of the line profile parameters when the
star is fading and compared it with the observed one.

\begin{figure*}
  \centering
  \includegraphics[width=14cm]{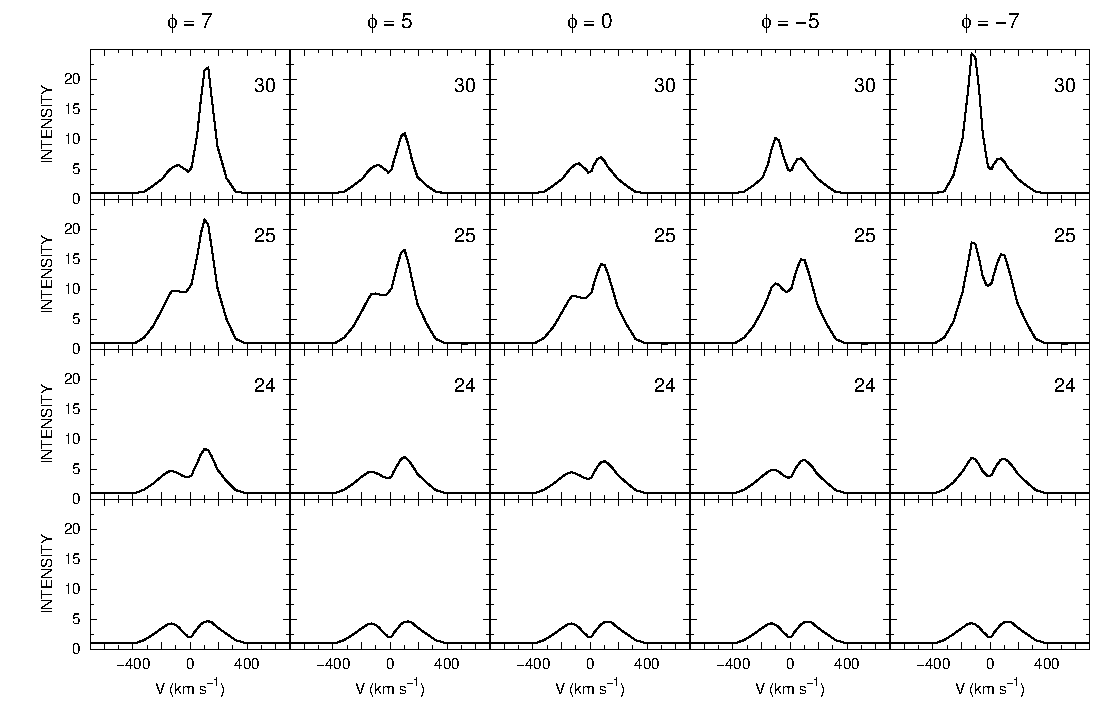}
\caption{The $H_\alpha$ line profiles for the hybrid model H1
  during an eclipse for the case of the screen width of 20$R_*$. The
  line profiles changed from the bright state (the lowest row) up
  to the total eclipse (the top row). Each column refers to
  different values of the azimuthal angle $\phi$ ($\phi$=0 means a
  central eclipse). The numbers in the plots refer to the screen
  heights expressed in stellar radii. See details in the text.}
\label{pro1}
\end{figure*}
\begin{figure*}
  \centering
     \includegraphics[width=14cm]{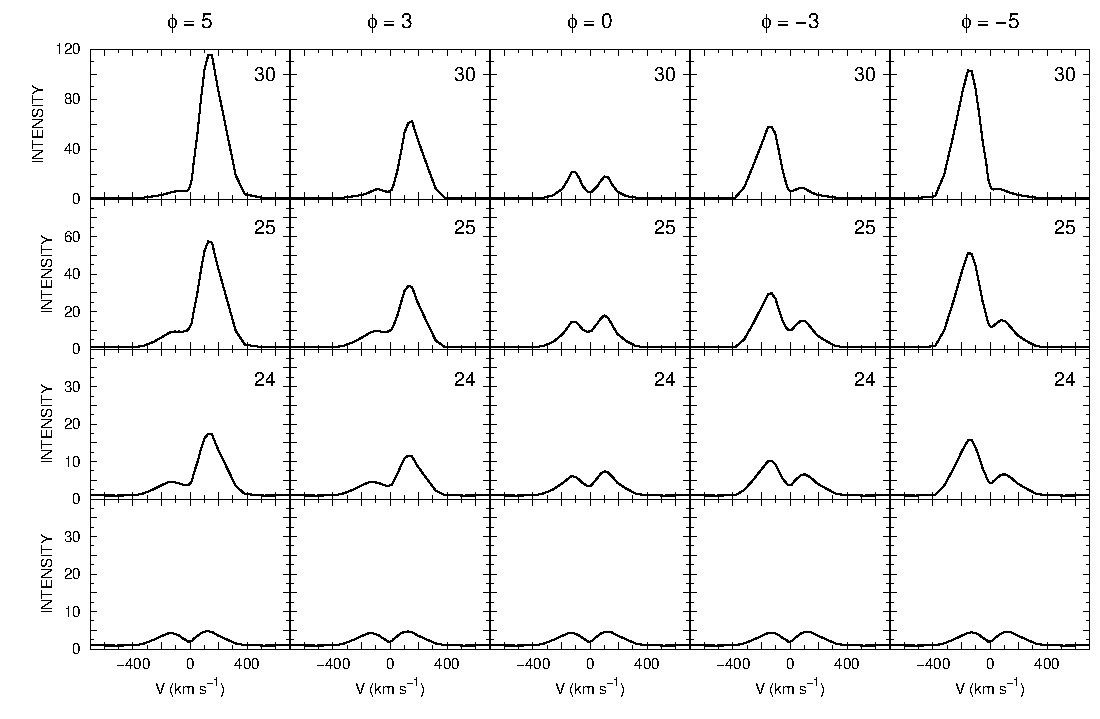}
  \caption{The same as in Fig.~\ref{pro1} but for the case of the screen
  width of 10$R_*$.}
  \label{pro2}
\end{figure*}
\begin{figure*}
  \begin{center}
  \centering
\includegraphics[width=14cm]{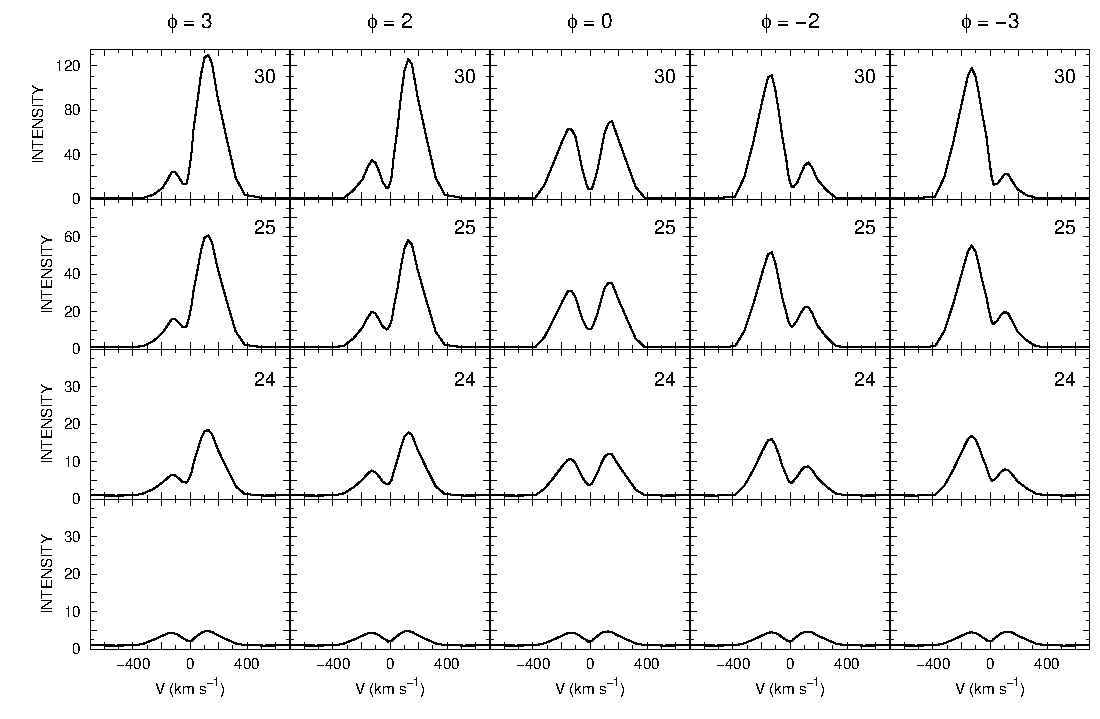}
  \caption{The same as in Fig.~\ref{pro1} but for the case of the screen
    width of 5$R_*$.}
  \label{pro3}
  \end{center}
\end{figure*}

Figures~\ref{pro1}, \ref{pro2}, and \ref{pro3} show the H$\alpha$
line profiles calculated with model H1 for different widths of the
screen equal to 20, 10 and 5$R_*$ respectively. The columns refer
to different values of the azimuthal angle $\phi$, and the rows
show the line profiles at different heights of the screen. The
lowest rows demonstrate the line profiles when the star is in the
normal (bright) state. When the height of the screen increases,
the star continuum decreases, and the intensities of the line
profiles also decrease. In order to compare the model profiles
with those observed, we normalized their intensities to their
current continuum, thus the intensities of the model profiles grow
with decreasing brightness. When the dusty screen reaches the
height of $\sim 30R_*$ the star continuum falls off so strongly
that only the scattered light dominates.

When the star enters the brightness minimum the violet peak gradually
decreases if the screen is located at positive values of
$\phi$ because it covers the matter approaching towards the observer.
When the star goes out of the minimum the blue peak is restored and the
red one decreases because the screen now covers the matter moving away
from the observer (negative $\phi$). This is seen more clearly in the case
of a small screen (Figs.~\ref{pro2} and \ref{pro3}).

\begin{figure*}
\begin{centering}
\includegraphics[width=5.5cm]{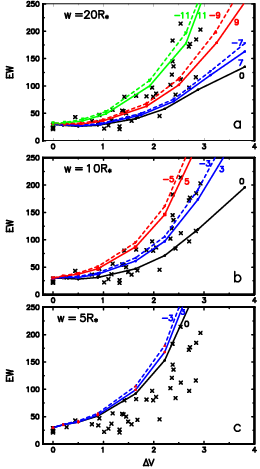}
\includegraphics[width=5.5cm]{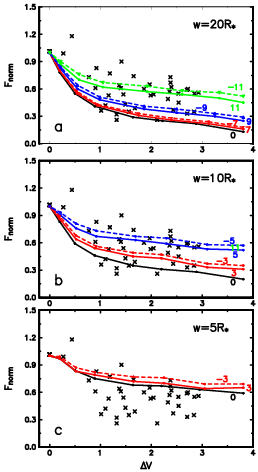}
\includegraphics[width=5.5cm]{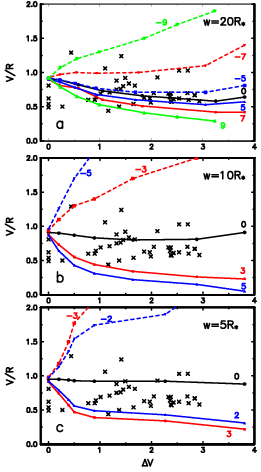}
\caption{Behavior of the equivalent width (left), flux in the
H$\alpha$ line (medium), and ratio of the blue to red intensity
peaks (right) with the brightness of the star during eclipses in
the hybrid model H1. The numbers mark values of the azimuthal
angle $\phi$. The crosses mark the observed data points.}
\label{ew1}
\end{centering}
\end{figure*}

The behavior of the equivalent widths (EWs), the ratio of the blue
to red intensity peaks of the H$\alpha$ line profiles (V/R), and
the radiation fluxes in the line during eclipses on the star
brightness in the visible region of the spectrum for the hybrid
model H1 is shown in Fig.~\ref{ew1}. The crosses mark the
observational data.

The EWs, the normalized $F_{\alpha i}$, and the V/R peak ratios
obtained for the screen of 20$R_*$ width (Fig.~\ref{ew1}a) are in
a good agreement with the observed parameters in a large range of
the azimuthal angles ($\pm 11^\circ$).
For the V/R intensity peak ratios, however, the range of suitable values
of $\phi$ is slightly less. This may be connected with the fact that
the initial line profile does not reproduce the observed profile
perfectly, for example, the deep central gap. Nevertheless, this
screen is the only (from those considered) that demonstrates a
coincidence of all parameters with the observed ones. Calculations
of eclipses with the smaller screen (10~$R_*$) gives acceptable results
for the twice narrower $\phi$ range ($\pm 5^\circ$), and reproduces
the shapes of the line profiles (V/R) only near the central eclipse
($\phi \approx 0$), see Fig.~\ref{ew1}b. The screen of a width of
5$R_*$ is not able to reproduce the observed line profile parameters
(except for the very central eclipse with $\phi=0^\circ$).

\begin{figure*}
\begin{centering}
\includegraphics[width=5.5cm]{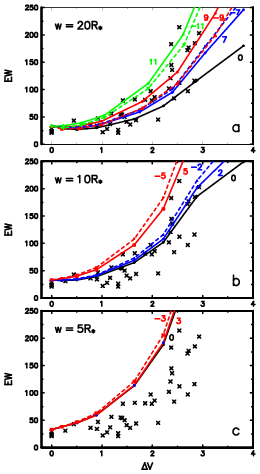}
\includegraphics[width=5.5cm]{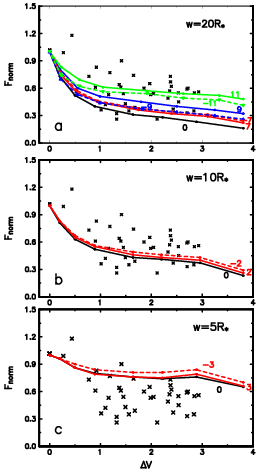}
\includegraphics[width=5.5cm]{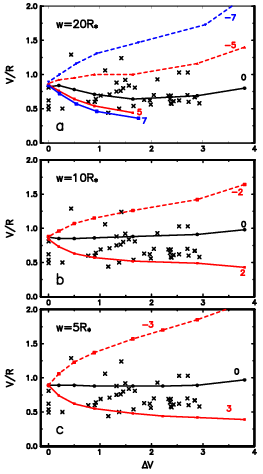}
\caption{The same as in Fig.~\ref{ew1} but for the hybrid model H2.}
\label{ew2}
\end{centering}
\end{figure*}

The same procedure has been followed for the hybrid model H2. The
H$\alpha$ line profiles for the cases of 10$R_*$ and 20$R_*$ widths of
the screen is present in the Appendix (Sect.~\ref{appendix-a}).
It should be noted that here the disc wind is launched from a more extended
region of the disc and is "more flared". One can see from
Figs.~\ref{pro4} and \ref{pro5} that for both values of the screen
widths the line profiles are practically single and blue or red
shifted\footnote{Such
  a blue shifted profile was observed for UX~Ori in the deep minima
  \citep{gri94} but has so far not been observed for RR~Tau.}.
As shown in Fig.~\ref{ew2}, for this model the suitable width of
the screen is 20$R_*$. The shorter V/R tracks in Fig.~\ref{ew2}a
for angles $5^\circ$ and 7$^\circ$ means that the line profiles do
not have the blue peak at the current height of the screen
(Fig.~\ref{pro4}). In other words, the line profile is an asymmetric
and single one.

\subsection{The second obscuration scenario}

First we estimated an optimal width of the obscuring body using a
typical duration of an eclipse. If we assume that the latter lasts
a week and keep in mind the Keplerian velocity at the dust
sublimation radius, then the width of the screen will be about of
20~$R_*$. Suitable parameters of the screen's opacity are given
with the Gauss function (eq. 2) with $\sigma = 2$ and $\tau_0 =
4$.

\begin{figure}
    \centering
\includegraphics[width=3.3cm]{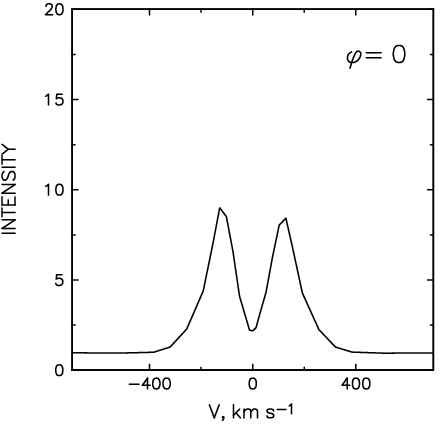}\\
\includegraphics[width=6cm]{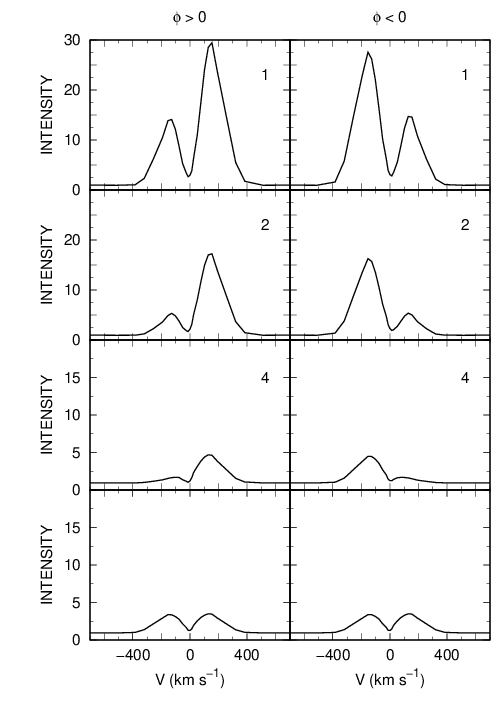}
\caption{The H$\alpha$ line profiles in the H1 hybrid model in the
second obscuration scenario. Numbers mark values of the azimuthal
angles $\phi$. The lower row presents line profile calculated at
the large values of $\phi$. Details are in the text.}
    \label{pro0}
\end{figure}

The changes in the shape of the H$\alpha$ line profile in the
second obscuration scenario are present in Fig.~\ref{pro0}. The
results of the calculations are valid for both screen widths
$w_s=10$ and $20R_*$. The left column demonstrates the line
profile from the bright state of the star (the lowest row) when
the screen is located at a large positive $\phi$. As the screen
moves towards the center it obscures the part of the envelope
approaching towards an observer until it reaches the center where
$\phi$ and the impact parameter $p$ are equal to zero.

During this passage the blue peak of the line profile "fades", and
the red one increases relatively to the weakening continuum. At
$\phi=0^\circ$ the screen covers the star at the densest part of
the envelope (practically symmetrically) that results in the weak
line profile with approximately equal peaks. As the screen
continues to move towards negative $\phi$, it obscures the
recessing part of the envelope, and the line profile shape is
inverted, compared to that at positive azimuthal angles. It should
be noted that such a behaviour of the H$\alpha$ line profile shape
during the RR Tau eclipse was not observed.

\begin{figure}
    \centering
\includegraphics[width=5cm]{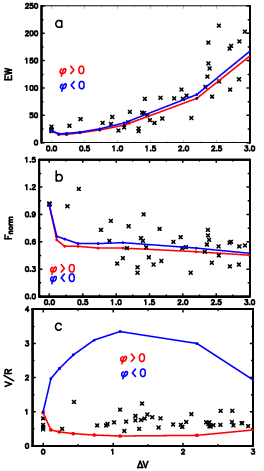}
\caption{Behaviour of the equivalent width (top), the flux in the
H$\alpha$ line (medium), and the ratio of the blue to red
intensity peaks (bottom) with the brightness of the star during an
eclipse in the model H1. The second obscuration scenario.}
    \label{pro01}
\end{figure}
Figure~\ref{pro01} shows calculated EWs, V/R ratios and radiation
fluxes in the H$\alpha$ line during eclipses in the framework of
the second obscuration scenario for the hybrid model H1.
In Fig.~\ref{pro01} a and b one can see that in this case
the equivalent widths of the profiles and the radiation
fluxes in the line are in a good agreement with the observations.
The peak ratios, however, contradict to the observational data both
at positive and negative azimuthal angles, see Fig.~\ref{pro01} c.

The disc wind model DW2 in the hybrid model H2 differs from the
disc wind model DW1 with a more extended launching region,
a larger first opening angle, and a greater mass loss rate.
From Fig.~\ref{ew2} representing changes in the main line profile
parameters with the star brightness in the framework of the first
obscuration scenario, one can see that the screen with the small
width (5$R_*$) does not give an agreement between the calculated
and observed equivalent widths. As for other screens' widths,
conclusions are the same as those made for model H1: the gas
and dust screen with the width of 20$R_*$ and more is preferable
because it gives a wider range of azimuthal angles where the line
profile parameters agree with the observed ones. However, the
calculated H$\alpha$ line profiles become strongly asymmetric and
single with the screen lifting, except for the central eclipse
(see Figs.~\ref{pro4},~\ref{pro5}), and that was not observed.

Figure~\ref{pro7} presents calculated EWs (top), radiation fluxes
in the H$\alpha$ line (medium) and V/R ratios (bottom) during
eclipses in the framework of the second obscuration scenario for
the hybrid model H2. The results are valid for screens with widths
equal to 10 and 20$R_*$. It is seen from the figure that an agreement
with observations exists only for radiation fluxes in the H$\alpha$
line (Fig.~\ref{pro7}b) and the peak ratio at the positive azimuthal
angles $\phi$. For other line profile
parameters we obtain a discrepancy between modelled and
observational data.

It should be noted that the parameters of UXORs are constrained in
rather narrow ranges. Therefore, the models calculated above may also
be used for other stars of this family.

\section{Discussion}
\label{sec:discussion}
In UX~Ori stars, as well as in Herbig~AeBe stars, the formation of the
most intensive emission line, H$\alpha$, is mainly due to
an extended disc wind, while in the formation of other Balmer lines, such
as H$\beta$ and H$\gamma$, the role of the magnetospheric accretion noticeably
increases because these lines originate in the less extended envelope
near the star. However, modelling the hydrogen line profiles for RR~Tau showed
that the disc wind plays a dominant role also for the H$\beta$ line.
Therefore we discuss the properties of the disc wind in more
details. Our modelling permits us to accentuate and discuss the following
conclusions.

\subsection{The disc wind parameters.}
The best coincidence of the calculated H$\alpha$ line profiles with those
observed with the NOT at the bright state of RR~Tau (i.e. the shape and
intensities of the line peaks) are as follows: (1) the wind launching region
has to begin not very far from the star. The best values for the inner
cylindric radii $w_1$ are 2 - 3~$R_*$.
The optimal values of the mass loss rate in such a case are ranged
from $(2 - 5)10^{-9}M_\odot$ yr$^{-1}$. For more distant launching
regions of the disc wind, the mass loss rate has to be larger in order to
fit the observed peak intensities. In this case the central gap of the line
profile will be filled in with additional emission and the shape of the
modelled profiles will deviate from the observed one.

We also calculated the hybrid model H1 conserving accretion and wind
parameters but using the values of the mass and radius of RR Tau recently
published by \citet{guz21}. The shapes of the line profiles with parameters
of the star given in the present paper and those by \citet{guz21} coincide;
the difference is in the value of the mass loss rate: in our case it is
$3\cdot10^{-9}M_\odot$ yr$^{-1}$, while in the case of \citet{guz21} it is
$5\cdot10^{-9}M_\odot$ yr$^{-1}$. Also in the latter case, the dust
sublimation radius is smaller ($\sim 33~R_*$).
The dynamics of the line profiles during eclipses does not change.

\subsection{The obscuration scenario and properties of the
obscuring body.}
Investigation of the transformation of the H$\alpha$ line profiles
revealed that the first obscuration scenario, where the vertical motion of
the dust screen dominates, is able to explain well all the variations of the
line parameters during eclipses, namely, the equivalent widths, the
radiation fluxes in the line, and the ratio of the line peaks. The
optimal size of the width of the screen has to be $ \geq 20~R_*$.

The reasons for the observed brightness minima are not established; they may
have a different nature. The properties and geometry of the dust screen
in the discs of young stars, and the behaviour of the linear polarization
and brightness during eclipses with such a screen in UXORs, has been recently
considered by \citet{shu19a,shu19b}.
They showed that the large scale perturbations in the discs can strongly
influence the photometric and polarimetric variability of UXORs. Since the
observed eclipses require large vertical matter perturbations, it was
suggested that the latter may be connected with
cyclonic vortices capable of lifting the dust above the disc surface
\citep[e.g.][]{god00,bar03}, or with an azimuthally structured
dusty disc wind \citep{tam08}. Also the charged dust grains can
rise high above the disc due to their interaction with the magnetic
field \citep{tur14}. \citet{giac19} showed that
the 0.1 - 1 $\mu$m crystalline silicate dust in the outer layers of
the protoplanetary discs of many objects (the Herbig stars are
among them), may be carried away by the different types of disc winds.
This dust is lifted above the disc surface and returns back at large
distances from the star.

\subsection{Variability of the other lines}
\subsubsection{The He I 5876 \AA~ line }

In the deep brightness minima, when the dust screen completely obscures the
stellar disc, one can see a strong transformation of the spectral lines
forming in the close vicinity of the star.
One such line is the He~I 5876 \AA \ line. It is absent in the spectra of
normal A stars on the Main Sequence. In the spectra of the young
Herbig~Ae stars this line forms in the magnetospheres of the stars in
the infalling gas stream. Because of the rapid rotation of UXORs, their
magnetosphere has a small size:  1.5 - 2.5 stellar radii. For this
reason the dusty screen obscuring the star can also cover completely the
region of formation of the helium absorption line. As a result, this line
disappears. This is clearly observed when one compares the spectra of
RR~Tau obtained in a bright state versus in deep minimum, see
Fig.~\ref{fig:obs1}, in the panel of the sodium lines, date 06.03.2019.

\subsubsection{The sodium D Na I lines}

These lines originate (at least partially) in the magnetosphere of
the star. This is demonstrated by the red-shifted absorption components of
the Na~I D lines which are seen in the spectra obtained during the bright
state of the star, and absent during brightness minima. The reasoning is the
same as that in the case of the He lines: an obscuration of the stellar disc
with the dusty screen. In the brightness minima the emission components of
the lines are slightly shifted to the blue side. During the bright state
they are faintly noticed. The source of the emission may be the peripheral
regions of the disc wind. This is indicated by a weak emission asymmetry.
The central narrow absorbtion in the sodium doublet lines is formed in the
circumstellar and interstellar medium.

\subsubsection{The Fe II 4924 \AA~ line}

In the bright state of RR Tau the Fe~II 4924 \AA~ line is observed as a pure
photospheric line. The two other components of this triplet look analogously.
With the brightness attenuation an emission appears from the blue side of the
absorption line; it strengthens and gradually completely blends the absorption line.
A similar transformation of the Fe~II triplet lines with RR~Tau fading was observed
by \citet{rod02} in spectra with a lower resolution. The authors suggested that
the emission components of the Fe~II triplet have a wind origin.

Our observations confirm this result. The high resolution of our spectra permits us
to determine the typical velocity of the gas motion in the region of the emission
formation: it is $\approx$ 200 km s$^{-1}$. Its significant part refers to the
poloidal component; this results in an observed shift of the emission
towards the blue side. During the bright state of the star this emission
is weak, but present, and it influences the position of the measured line center
by shifting it to the red. Hence an important conclusion: the radial velocities
of young stars determined with the help of photospheric lines may have systematic
errors because of the blending the photospheric lines with the emission lines of
the disc winds.

\subsubsection{The H$\beta$ line}
As shown in Fig.~\ref{fig:obs1} this line demonstrates an inverse P~Cygni
line profile during the bright state, and a double peak profile during brightness
minima. The peak ratio V/R can be slightly greater or less than one. Similarly
to the H$\alpha$ line profiles, the intensity of the H$\beta$ line increases during
an eclipse. As mentioned above, the H$\beta$ line responds much stronger to the
variable processes in the star's vicinity, in particular to irregular and strong
changes in the mass accretion rates, as shown in Fig.~\ref{hb19} and Fig.~\ref{hb22}.
For example, one can see from Fig.~\ref{hb19}, that
the line profile can radically change in one night although the brightness of the star
did not change. Nevertheless, the red peak of the line profile (red line) practically
disappears.

Another interesting example is shown in Fig.~\ref{hb22}. Here one can see transformation
of the shape of the observed line profile (black) over five nights. It is easy to
notice by comparing it with the computed H$\beta$ line profile for one of the model
(blue), that is the same at each plot. In spite of the fact that the brightness remained
nearly the same (about V $\approx$ 11 mag), the observed profile modifies the shape and
at last coincides with the computed one. Such a strong and rapid variability of this line
makes its modelling more unpredictable than in the case of the H$\alpha$ line.

\subsubsection{[O~I] 6300 \AA~ line}
This forbidden line (as well as the [O~I] 6363 \AA~ line) is observed in all spectra of
RR~Tau. Its equivalent width increases when the star fades, and this dependence is
described well by the relation: $V = 12.1 + 2.5 \log(EW_{[OI]6300})$. It means that the
flux in this line did not change during our observations, in agreement with the variable
CS extinction model. A similar relation for this line was obtained by \citet{rod02}.

According to our measurements the fitted Gaussian FWHM of the line is around
1.1 $\pm$ 0.1 \AA \ or translating to line widths $\Delta v$~=~53 $\pm$ 5~km~s$^{-1}$.
We find that the heliocentric RV of the 6300 \AA \ line is 6 $\pm$ 3~km~s$^{-1}$ and that
of the 6363 \AA \ line is -2~$\pm$~2~km~s$^{-1}$. With the stellar RV of 11 km s$^{-1}$
\citep{grik01}, this means the [O~I] lines are slightly blue-shifted. Both of these
measurements suggest that the photoevaporated disc wind \citep{er10} is the most
probable location of the [O~I] line formation.

\subsubsection{DIB 6283 \AA}

The diffuse interstellar band (DIB) 6283 \AA \ was observed in all
the spectra of RR~Tau, and we did not find any changes in its
parameters during the brightness minima. This means that the CS
extinction does not add anything to the DIB and that it has a purely
interstellar origin.

\subsubsection{Comparison with the previous observations with NOT}
Comparing the observational results of the seasons from 2019 - 2022
with previous ones from 1995 - 1996 obtained also with the NOT
\citep{grik01}, one can see that the H$\alpha$ line always shows the
double peak line profile but with the peak ratios blue-to-red
(V/R) both less than one and greater than one. The H$\beta$ line
profiles during the bright state has both the inverse P~Cygni and
double peak line profiles. It should be noted that in the above
cited paper the spectra of RR~Tau were obtained during the bright state
except for one night when the star noticeably weakened. In all the
spectra of RR~Tau the central narrow absorption in the Na~I D lines was
slightly deeper in the 1995 - 1996 data compared to the
recent observations presented in this paper. Since the spectral
resolution in both programs was the same, this difference means that
the conditions for propagation of the stellar radiation to an observer
in the Na~I~D resonance line frequencies could slowly change with time.

\section{Conclusion}
Our main conclusions are as follows:

1. The results of our spectroscopic monitoring of RR~Tau with FIES at the NOT
confirm the results of the previous observations of this star by \citet{rod02}
obtained with a low resolution spectrograph.

2. The higher spectral resolution (R = 25000) permitted us to
investigate the spectral variability of RR~Tau in more detail.
Modelling the H$\alpha$ line observed in the bright state and
during the deep minima shows that the reason for the eclipses
in this star is the vertical rise of the dust above the disc.
The heterogeneous dusty disc wind can, in principle, provide
such a scenario.

3. During the deep minima we observed the appearance of weak
blue-shifted emission in some metallic lines (Fe~II, Na~I~D and some
others). Their radial velocities suggest the disc wind as the
most probable source of this emission. These emission components
are not observed during the bright state of the star, but their
hidden existence will affect the photospheric line profiles and
shift the measured stellar radial velocity towards the red. This
effect has to be present in all young stars with intensive disc winds.

4. The constant luminosity of the forbidden lines [O~I] 6300/6363
  \AA \ during the eclipses means that these lines originate beyond the
  screen. The link between the EWs of these lines with an amplitude of
  the stellar brightness definitely indicates that the circumstellar
  extinction is the main reason for the brightness variability of the
  star. These lines are
narrow and slightly blue-shifted. The photoevaporated disc wind seems
to be the best place of their formation.

It should be stressed that these conclusions are made for one UXOR
star - RR~Tau. In order to obtain a more general picture we need
spectroscopy with high spectral resolution of other UXORs at their
different brightness levels.

\section*{Acknowledgements}
\addcontentsline{toc}{section}{Acknowledgements}
  We thank the referee for his/her useful comments and suggestions which helped to improve the paper.
  V.P.G. and L.V.T. would like to acknowledge the support of the Ministry of Science and
  Higher Education of the Russian Federation under the grant 075-15-2020-780 (N13.1902.21.0039).
  A.A.D. thanks staff and students at the NOT for the continuous support, making the flexible
  scheduling possible and efficient.

\section*{Data availability}

All the FIES data used in this paper is available in the online FITS archive\footnote{\url{http://www.not.iac.es/observing/forms/fitsarchive/}. } of the Nordic Optical Telescope, raw data as well calibration data. The FIEStool pipeline reduced spectra are available by contacting staff at the NOT. Practically all the photometry observed by SAAF are available in AAVSO\footnote{\url{https://www.aavso.org/}}, or otherwise on request to the authors.

\bibliographystyle{mnras}
\bibliography{rrtau}

\appendix
\section{Results of calculation of the hybrid model H2}
\label{appendix-a}

\begin{figure}
    \centering
    \includegraphics[width=6cm]{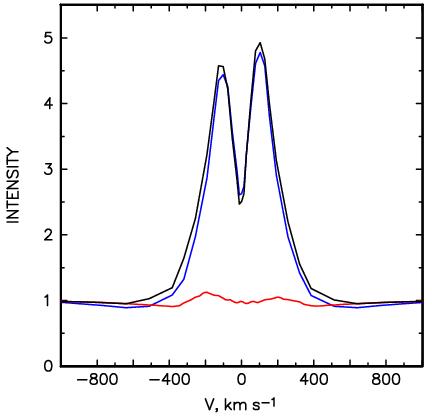}
\caption{Contribution of the different emitting regions to the
total H$\alpha$ line profile in the hybrid model H1: the accretion
region (red), the disc wind region (blue).} \label{rm4}
\end{figure}

\begin{figure*}
    \centering

\includegraphics[width=14cm]{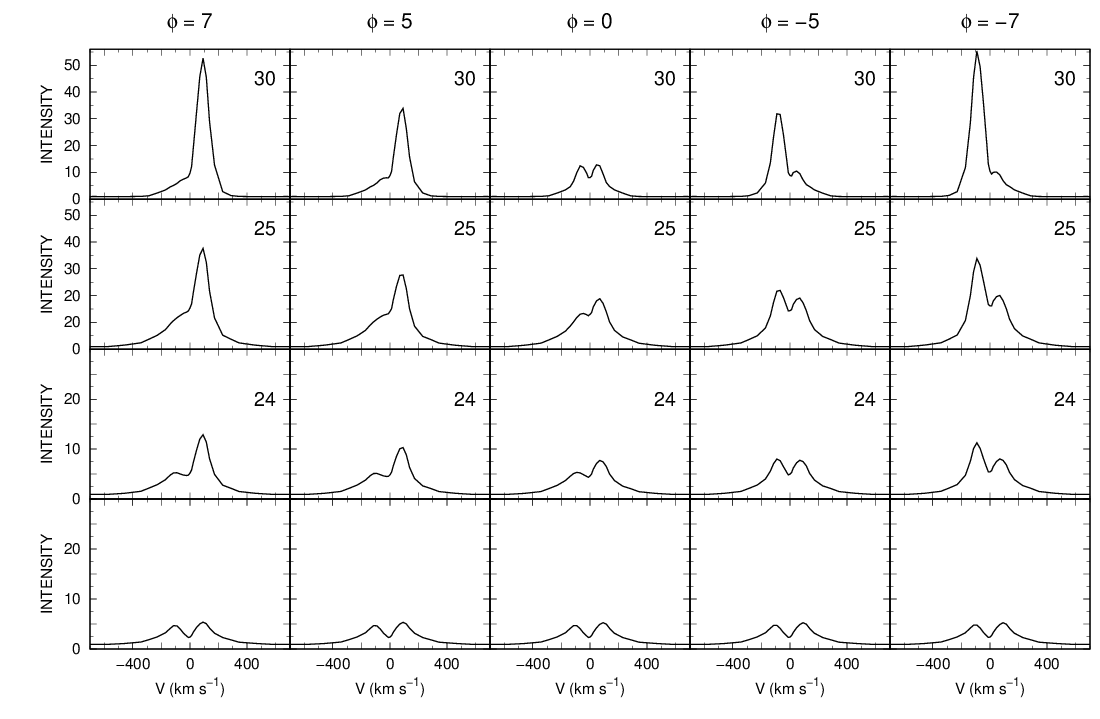}
\caption{The H$\alpha$ line profiles for the hybrid model H2
during an eclipse for the case of the screen width of 20~$R_*$.
The line profiles change from the bright state (the lowest row) up
to the total eclipse (the top row). The numbers at the plots refer
to the screen heights expressed in stellar radii.} \label{pro4}
\end{figure*}
\begin{figure*}
    \centering
\includegraphics[width=14cm]{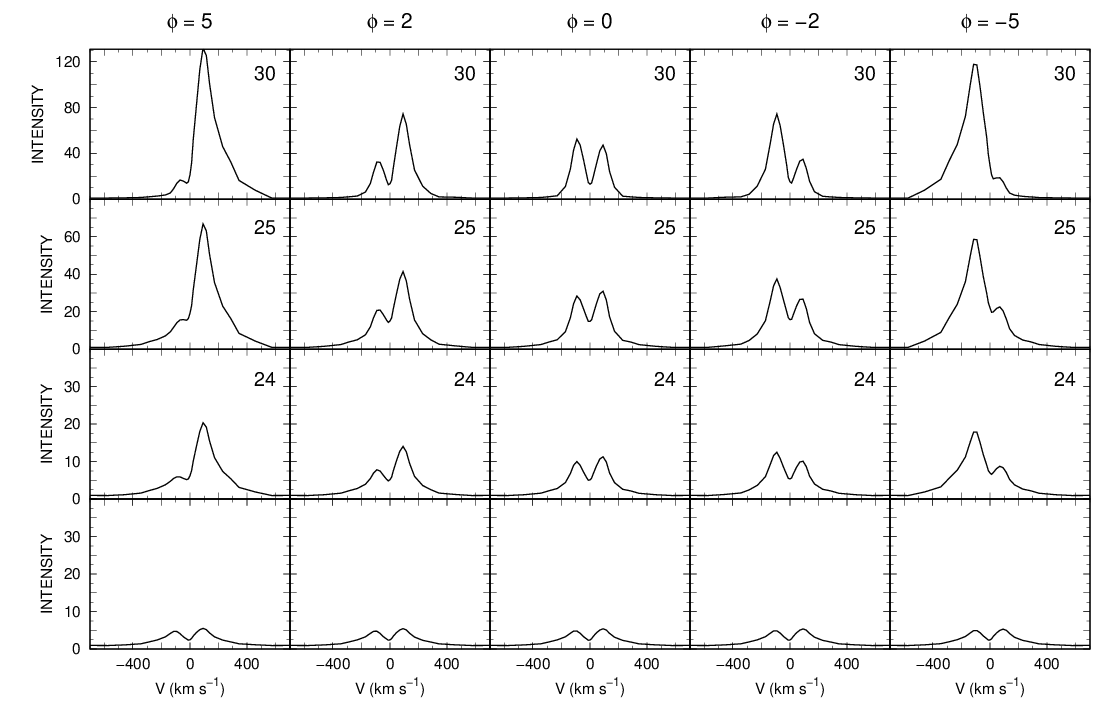}
  \caption{The same as in Fig.~\ref{pro4} but for the case of a screen
  width 10~$R_*$.}
\label{pro5}
\end{figure*}

\begin{figure}
    \centering
\includegraphics[width=5cm]{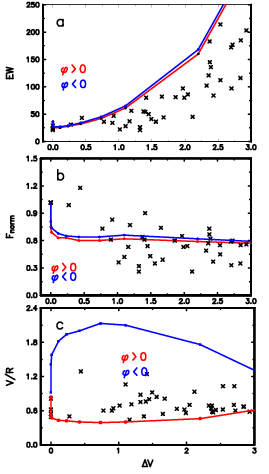}

\caption{Hybrid model H2. The behavior of the equivalent width
(top), the flux in the H$\alpha$ line (medium), and the ratio of
the blue to red intensity peaks (bottom) with the brightness of
the star during eclipses. The second scenario. Parameters used in
Eq. 2 are $\sigma = 3, \tau = 4$.}
    \label{pro7}
\end{figure}

\section{H$\beta$ line}
\label{appendix-b}

\begin{figure}
    \centering
\includegraphics[width=6cm]{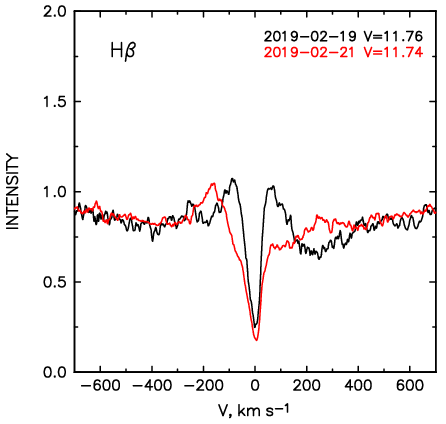}

\caption{A sharp change in the H$\beta$ line profile during two
  nights in 2019 with the same brightness in the V-band.
  The colours of the line profiles correspond to the same colors
  of dates and brightness values.}
    \label{hb19}
\end{figure}

\begin{figure}
    \centering
\includegraphics[width=5cm]{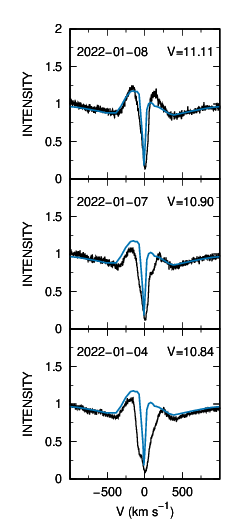}

\caption{The rapid change in the observed H$\beta$ line profile
  obtained with the NOT over a time span of five nights in 2022
  (black). The dates are shown in the plots. For comparison we give
  the computed H$\beta$ line profile which is the same at each plot
  (blue). The blue profile is obtained in the framework of the
  accretion model with parameters given above and the disc wind
  models with the following parameters: the mass loss rate is
  2$\cdot 10^{-9}M_\odot$ yr$^{-1}$, the first half open angle is
  45$^\circ$, the disc wind starts from 3~$R_*$ from the star.}
    \label{hb22}
\end{figure}

% Don't change these lines
\bsp    % typesetting comment
\label{lastpage}
\end{document}